\documentclass[twocolumn,twocolappendix]{aastex631}

\newcommand{\gcc}{\,g\,cm$^{-3}$} % gram per cm-cubed
\newcommand{\ergs}{\,erg\,s$^{-1}$}	% erg per s
\newcommand{\ms}{\,M$_\odot$} %  mass

\usepackage{amsmath}
\usepackage{newtxtext}
\usepackage{newtxmath}
\usepackage{graphicx}
\graphicspath{{./}{figures/}}

\shorttitle{$R$-process Nucleosynthesis of Subminimal Neutron Star Explosions}
\shortauthors{Yip et al.}

\begin{document}
\title{$R$-process Nucleosynthesis of Subminimal Neutron Star Explosions}

\author[0000-0002-3311-5387]{Chun-Ming Yip}
\affiliation{Department of Physics and Institute of Theoretical Physics, The Chinese University of Hong Kong, Shatin, N.T., Hong Kong S.A.R., People's Republic of China}

\author[0000-0002-1971-0403]{Ming-Chung Chu}
\affiliation{Department of Physics and Institute of Theoretical Physics, The Chinese University of Hong Kong, Shatin, N.T., Hong Kong S.A.R., People's Republic of China}

\author[0000-0002-4972-3803]{Shing-Chi Leung}
\affiliation{Department of Mathematics and Physics, SUNY Polytechnic Institute, 100 Seymour Road, Utica, New York 13502, USA}

\author[0000-0002-4638-5044]{Lap-Ming Lin}
\affiliation{Department of Physics and Institute of Theoretical Physics, The Chinese University of Hong Kong, Shatin, N.T., Hong Kong S.A.R., People's Republic of China}

\begin{abstract}

We show that a minimum-mass neutron star undergoes delayed explosion after mass removal from its surface. 
We couple the Newtonian hydrodynamics to a nuclear reaction network of $\sim4500$ isotopes to study the nucleosynthesis and neutrino emission during the explosion. 
An electron antineutrino burst with a peak luminosity of $\sim3\times10^{50}$\ergs\, is emitted while the ejecta is heated to $\sim10^{9}$\,K. 
A robust $r$-process nucleosynthesis is realized in the ejecta. 
Lanthanides and heavy elements near the second and third $r$-process peaks are synthesized as end products of nucleosynthesis, suggesting that subminimal neutron star explosions could be an important source of solar chemical elements.

\end{abstract}


%----------------------------------------------------------------------
\section{Introduction}

Neutron star binaries have been intensively investigated for several decades, both observationally and theoretically. 
The discovery of pulsar PSR B1913+16 in a neutron star binary by \citet{1975ApJ...195L..51H}, and its orbital decay due to gravitational wave radiation \citep[][]{1981SciAm.245d..74W} revealed the evolution of close neutron star binaries. 
The first-ever multi-messenger detection of a binary neutron star merger in 2017 \citep[][]{Abbott_2017,Goldstein_2017,Savchenko_2017} confirmed the encounter of neutron stars in close binaries. 
Apart from direct merger, \citet{1977ApJ...215..311C} discussed the mass transfer in a close neutron star binary with different initial masses. 
It was found that stable mass transfer can be established under certain circumstances, which would further enhance the asymmetry of the systems. 
\citet{1984SvAL...10..177B} subsequently derived that such mass transfer caused by Roche lobe overflow is slow and stable until the mass of the lighter and larger neutron star reaches $\sim0.15$\ms. 
After that, tidal disruption of the minimum-mass neutron star occurs with a characteristic timescale longer than its hydrodynamic timescale, so it evolves through a series of quasi-equilibrium states until reaching the minimum stable mass $\sim0.1$\ms\, allowed by the equation of state (EoS), and it explodes \citep[][]{1990AZh....67.1181B, Blinnikov_2021}. 
A recent postprocessing nucleosynthesis calculation by \citet{Panov_2020} revealed the production of heavy elements through this event.

The nuclear reactions involved in the expansion of neutron star matter were examined by \citet{1977ApJ...213..225L}. 
Simulations of subminimal neutron star explosions by \citet{1989ApJ...339..318C, 1991ApJ...369..422C, 1993ApJ...414..717C} showed the essential roles of nuclear reactions in the development of hydrodynamic instabilities. 
The timescale of weak interactions is considerably longer than the hydrodynamic timescale, and thus $\beta$-equilibrium cannot be always established dynamically when a neutron star is perturbed. 
The pressure of neutron star matter is not uniquely determined at a given density if the $\beta$-equilibrium is not achieved. 
Hence, the minimum mass of neutron stars determined from the EoS, in which the $\beta$-equilibrium condition is often assumed, does not necessarily correspond to the onset of instability. 
Hydrodynamic simulations of such a configuration with consideration of departure from $\beta$-equilibrium were performed. 
It was demonstrated that a sufficiently large ratio of mass removal, $\Delta S$, at the surface of the minimum-mass neutron star, can lead to an instability. 
\citet{1998A&A...334..159S} further presented the explosion of the subminimal neutron star in a simulation with an initial mass removal ratio $\Delta S=0.22$. 
An electron antineutrino burst with peak luminosity $\sim10^{52}$\ergs\, was predicted. 
Moreover, $r$-process was realized in the expanding neutron star matter when its density dropped rapidly, and rare-earth elements were produced.

Despite the inspiring predictions of neutrino burst and nucleosynthesis, several modifications of the model used by \citet{1998A&A...334..159S} are necessary for a more rigorous investigation. 
In this article, we revisit the reactive-hydrodynamic simulations of subminimal neutron star explosions with a more modern nuclear matter EoS. 
A nuclear reaction network, instead of the single nucleus approximation, is employed to compute the relevant nuclear reactions, including neutron capture, $\beta$-decay, and spontaneous fission, that are essential for nucleosynthesis. 
We predict the yield of stable elements produced through this event, and we show that lanthanides and heavy elements near the second and third $r$-process peaks are synthesized. 
Our results suggest that the explosions of subminimal neutron stars could be a potentially new class of astronomical $r$-process events in addition to other known heavy element production channels \citep[e.g.,][]{KAJINO2019109, 2020ApJ...900..179K}.

This paper is organized as follows: 
In Section~\ref{sec:Methodology}, we present the computational setup of the hydrodynamic simulations coupled with the nuclear reaction network. 
In Section~\ref{sec:Results}, we present the numerical results of the simulations and the nucleosynthesis calculations. 
We discuss and conclude the findings in Section~\ref{sec:Conclusions}.

%----------------------------------------------------------------------
\section{Methodology}
\label{sec:Methodology}

\subsection{Initial Configuration}

\begin{figure}[b!]
    %\plotone{figure1.png}
    \includegraphics[width=\columnwidth]{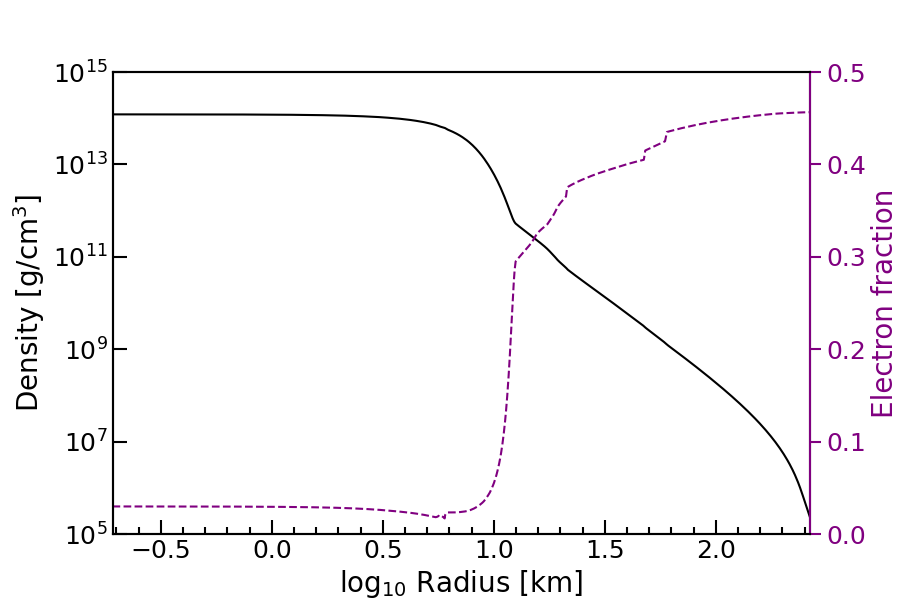}
    \caption{
    Initial configuration of the minimum-mass neutron star at hydrostatic and $\beta$-equilibrium for the EoS by Schneider, Roberts, and Ott \citep[SRO;][]{PhysRevC.96.065802,PhysRevC.100.025803} at uniform temperature \,$T=10^8$\,K. 
    The density (solid line) and electron fraction (dashed line) are plotted vs. the radial coordinates in logarithm scale.
    }
    \label{fig:initial configuration}
\end{figure}

We adopt the EoS by Schneider, Roberts, and Ott \citep[SRO;][]{PhysRevC.96.065802,PhysRevC.100.025803}, computed
using the NRAPR Skyrme parameterization \citep[][]{STEINER2005325}, an effective nonrelativistic Skyrme-type nuclear interaction model \citep[][]{LATTIMER1991331}, and nuclear statistical equilibrium (NSE) approximation at low densities and temperatures, available in \texttt{CompOSE}\footnote{\url{https://compose.obspm.fr}}, in our study. 
We construct the initial configuration of the minimum-mass neutron star with a uniform temperature \,$T=10^8$\,K under hydrostatic and $\beta$-equilibrium. 
A pure Newtonian gravity and spherical symmetry are assumed. 
The minimum-mass neutron star has a central density of about $1.211 \times 10^{14}$\gcc, a total mass $M$ of about $0.0896$\ms, and a radius $R$ of more than $250$\,km (see Figure~\ref{fig:initial configuration}). 
The general relativistic correction to the minimum-mass neutron star structure is negligible as its compactness $C = GM/Rc^2\sim5\times10^{-4}$ is very small.

\subsection{Hydrodynamic Simulation}

The set of Euler equations in the Lagrangian formalism, in spherical symmetry, are:
\begin{equation}
    \frac{\partial u}{\partial t} = -4 \pi r^2 \frac{\partial \left(p + q\right)}{\partial m} - \frac{G m}{r^2},
\end{equation}
\begin{equation}
    \frac{\partial r}{\partial t} = u,
\end{equation}
\begin{equation}
    \frac{1}{\rho} = 4 \pi r^2 \frac{\partial r}{\partial m},
\end{equation}
\begin{equation}
    \frac{\partial \varepsilon}{\partial t} = -\left(p + q\right) \frac{\partial \left( 1/ \rho \right)}{\partial t} + \dot\epsilon_{\text{nuc}} + \dot\epsilon_{\beta} + \dot\epsilon_{\nu\text{,}\beta} + \dot\epsilon_{f} + \dot\epsilon_{\nu\text{,}\text{th}},
\label{eq:energy equation}
\end{equation}
\begin{equation}
    q = \begin{cases}
        l^2 \rho \left( \frac{\partial u}{\partial r}\right)^2 & \text{if} \; \frac{\partial u}{\partial r} < 0, \\
        0 & \text{otherwise},
    \end{cases}
\end{equation}
where $G$ is the gravitational constant. 
The baryon mass $m$ contained within the radius $r$ from the center of the neutron star is chosen to be the Lagrangian mass coordinates, with the outward radial direction from the center defined as the positive direction. 
$u$, $\rho$, $p$, and $\varepsilon$ are the radial velocity, baryon mass density, pressure, and specific internal energy density of the fluid, respectively, as functions of time $t$ and $m$. 
A standard first-order Lagrangian finite-difference scheme \citep[][]{richtmyer1967difference} with $100$ uniform mass grid zones is employed to solve the Euler equations numerically with the introduction of an artificial viscosity term $q$ whenever compression occurs, where $l=1.0$\,$G$\ms/$c^2 = 1.477$\,km is the artificial viscosity coefficient having the dimension of a length. 
Explicit first-order time discretization subjected to the Courant–Friedrichs–Lewy (CFL) condition is chosen to update these hydrodynamic variables. 
In Equation~(\ref{eq:energy equation}), $\dot\epsilon_{\text{nuc}}$, $\dot\epsilon_{\beta}$, $\dot\epsilon_{\nu\text{,}\beta}$, $\dot\epsilon_{f}$, and $\dot\epsilon_{\nu\text{,}\text{th}}$ denote the extra heating/cooling rates of the mass elements associated with thermonuclear reactions, $\beta$-decays, $\beta$-decays neutrino emissions, spontaneous fission, and thermal neutrino emissions, respectively (see Section~\ref{sec:nuclear reaction network}).

\subsection{Nuclear Reaction Network}
\label{sec:nuclear reaction network}

\begin{figure}[t!]
    %\plotone{figure2.png}
    \includegraphics[width=\columnwidth]{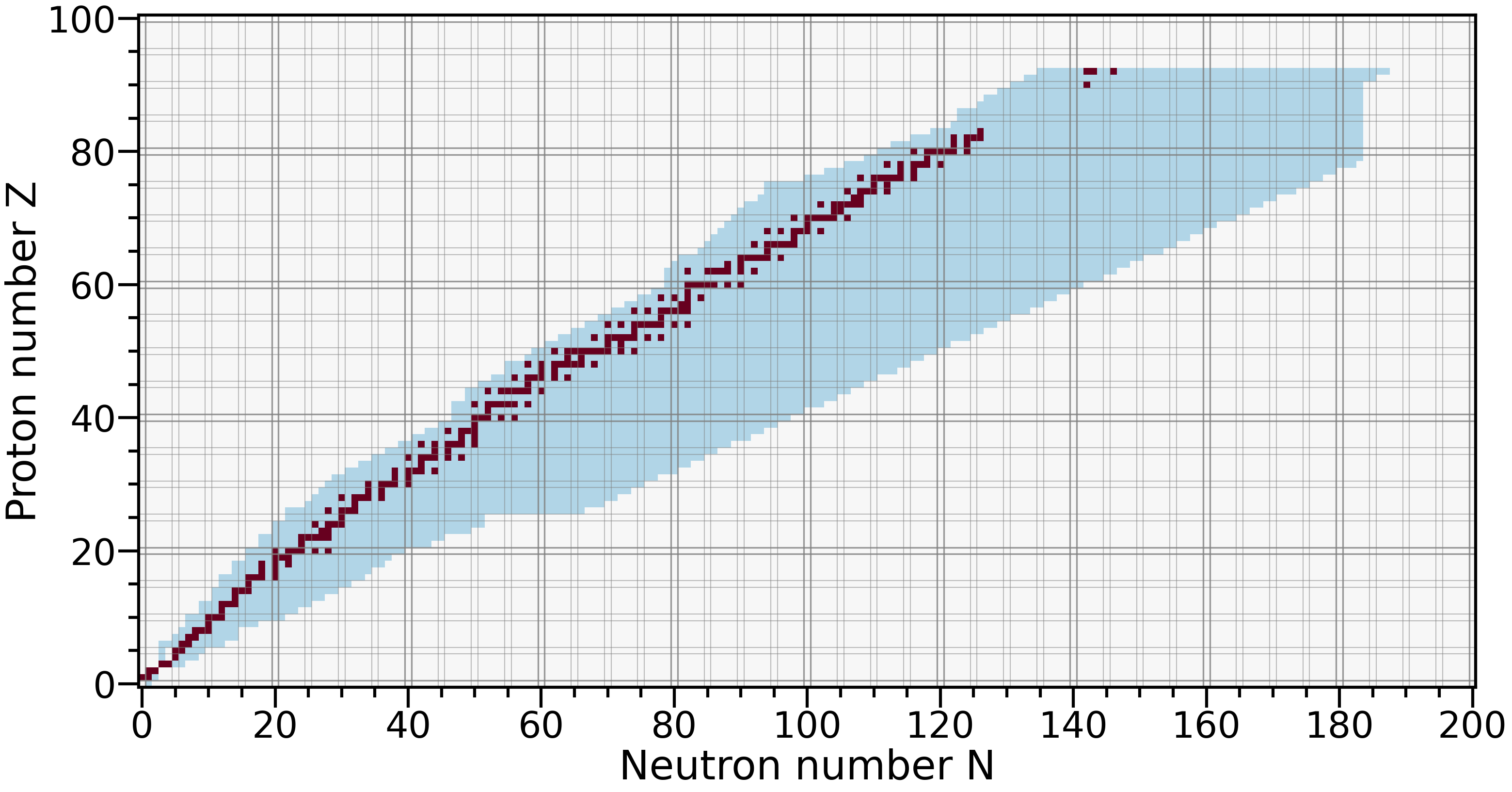}
    \caption{
    Isotopes included in the nuclear reaction network. 
    The colored region covers the isotopes considered in the network, and the grids in red further illustrate the stable isotopes in the solar system \citep[][]{lodders2019solar}. 
    Extremely neutron-rich isotopes with proton number $Z>80$ are not included because some of the nuclear reaction rates are not available to form linkages with other isotopes included.
    }
    \label{fig:network range}
\end{figure}

We implement and adopt the open-source code for nuclear reaction network computation developed by \citet{2000ApJS..129..377T}. 
A network consisting of $\sim4500$ isotopes (see Figure~\ref{fig:network range}) is chosen to study nucleosynthesis in this work. 
The atomic number $Z$ of the isotopes included in the network ranges from $Z_{\text{min}}=0$ for neutron to $Z_{\text{max}}=92$ for uranium. 
The stable elements synthesized through standard $r$-process channels and their short-lived neutron-rich isotopes temporarily produced during the $r$-process are included. 
The nuclear masses, partition functions, and thermonuclear reaction rates of pair reactions with proton, neutron, $\alpha$ particle, and photon, namely $(n,\gamma), (n,p), (p,\gamma), (\alpha,n), (\alpha,p), (\alpha,\gamma)$ processes, and their inverse processes, are used \citep[][]{2010ApJS..189..240C}. 
Some special reactions, such as deuterium fusion and carbon burning, are implemented additionally. 
The thermal energy generation/absorption rate due to the thermonuclear reactions $\dot\epsilon_{\text{nuc}}$ is computed by
\begin{equation}
    \dot\epsilon_{\text{nuc}} = - N_A \sum_i \frac{d Y_i}{d t} M_i,
\end{equation}
where $N_A$ is the Avogadro constant, and $Y_i$ ($M_i$) is the number fraction (nuclear mass) of isotope $i$. 
We sum over all isotopes $i$ in the network to find the net change in $Y_i$ for all isotopes after the thermonuclear reactions.

The $\beta$-decay half-lives and $\beta$-delayed neutron emission probabilities of nuclei provided by \citet{2003PhRvC..67e5802M} are adopted. 
The energy available $\Delta_j$ through the $\beta$-decay reaction $j$ of an isotope with nuclear mass $M(A,Z)$, where $A$ denotes the mass number of the isotope, is given by
\begin{equation}
    \Delta_j = \left[ M(A,Z) - M(A,Z+1) \right] - \mu_e.
\label{eq:beta decay}
\end{equation}
The chemical potential of electron $\mu_e$ enters the equation, since the Fermi level of degenerate electrons is high inside a neutron star. 
The endothermic $\beta$-decays are forbidden whenever $\mu_e$ is higher than the energy released. 
The average energy carried away by each electron antineutrino $\epsilon_{\nu_{\bar e}}$ can be expressed as \citep[][]{1998A&A...334..159S}
\begin{equation}
    \epsilon_{\nu_{\bar e},j} = \frac{3}{7} \Delta_j \frac{{\Delta_j}^2 + 7 \Delta_j \mu_e + 21 {\mu_e}^2}{{\Delta_j}^2 + 6 \Delta_j \mu_e + 15 {\mu_e}^2}.
\label{eq:neutrino energy}
\end{equation}
We take the approximation that the neutrinos produced through $\beta$-decays escape freely from the subminimal neutron star without any dissipation of energy \citep[][]{1989ApJ...343..254M}. Hence, the thermal energy generation rate $\dot\epsilon_\beta$ and the neutrino cooling rate $\dot\epsilon_{\nu,\beta}$ with respect to $\beta$-decays are computed by
\begin{equation}
    \begin{split}
    \dot\epsilon_\beta = N_A \sum_j \frac{d Y_j}{d t} \Delta_j, \\
    \dot\epsilon_{\nu,\beta} = -N_A \sum_j \frac{d Y_j}{d t} \epsilon_{\nu_{\bar e},j},
    \end{split}
\end{equation}
where summation over all $\beta$-decay reactions $j$, with $\Delta_j > 0$ under the given $\mu_e$ is performed. 
After the weak interactions, the heating effect owing to $\beta$-delayed neutron emissions is also calculated, which is found to be an insignificant contribution and is counted in $\dot\epsilon_{\text{nuc}}$ for convention.

The spontaneous fission half-lives of heavy nuclei $\tau_f$ (in a unit of yr) with atomic number $Z \geq 90$ in the network are evaluated using the semiempirical formula by \citet{2010NuPhA.832..220S}:
\begin{equation}
    \begin{aligned}
    \log_{10} \left( \tau_f \right) = & a \left( \frac{Z^2}{A} \right) + b \left( \frac{Z^2}{A} \right)^2 + \\ & c \left( \frac{N-Z}{N+Z} \right) + d \left( \frac{N-Z}{N+Z} \right)^2 + e,
    \end{aligned}
\end{equation}
where $N$ is the neutron number of the nuclei, and $a=-43.25203$, $b=0.49192$, $c=3674.3927$, $d=-9360.6$, and $e=580.75058$ are fitting parameters. 
We approximate the fission fragments after spontaneous fission by the equation
\begin{equation}
    \begin{aligned}
    \mathcal{N}_1 \left( A, Z \right) \longrightarrow & \mathcal{N}_2 \left( \gamma A, \gamma Z \right) + \\& \mathcal{N}_3 \left[ \left( 1-\gamma \right) A, \left( 1-\gamma \right) Z \right]
    \end{aligned}
\end{equation}
when a nucleus $\mathcal{N}_1$ undergoes spontaneous fission to release two daughter nuclei, $\mathcal{N}_2$ and $\mathcal{N}_3$. 
The fission fragment asymmetry parameter $\gamma$ controls the asymmetry of fission fragments. 
We set the values of $\gamma$ to be $0.50$, $0.45$, and $0.40$ in different models to investigate the influence of the fission fragment asymmetry on the subminimal neutron star explosions. 
The models with $\gamma = 0.50$ execute symmetric fission fragment approximation, while the other two classes of model roughly resemble the peaks of fission product yield distribution from more sophisticated fission fragment calculations \citep[e.g.,][]{Hao_2022}. 
After calculating the change in $Y_i$ of all isotopes $i$ in the network caused by spontaneous fission, the thermal energy generation rate through spontaneous fission $\dot\epsilon_{f}$ is computed by
\begin{equation}
    \dot\epsilon_{f} = - N_A \sum_i \frac{d Y_i}{d t} M_i.
\end{equation}

The thermal neutrino energy loss rate $\dot\epsilon_{\nu\text{,}\text{th}}$ through pair-, photo-, plasma-, bremsstrahlung, and recombination neutrino processes are calculated using the open-source subroutine for thermal neutrino emission\footnote{\url{http://cococubed.asu.edu/}}, which adopts the analytical fitting formulae by \citet{1996ApJS..102..411I}. 
It enters Equation~(\ref{eq:energy equation}) under the assumption that the thermal neutrinos escape freely from the neutron star. 
It is found that the cooling effect caused by thermal neutrino emission is negligible as the temperature reached by the exploding subminimal neutron star is not adequately high.

\begin{figure}[t!]
    %\plotone{figure3.png}
    \includegraphics[width=\columnwidth]{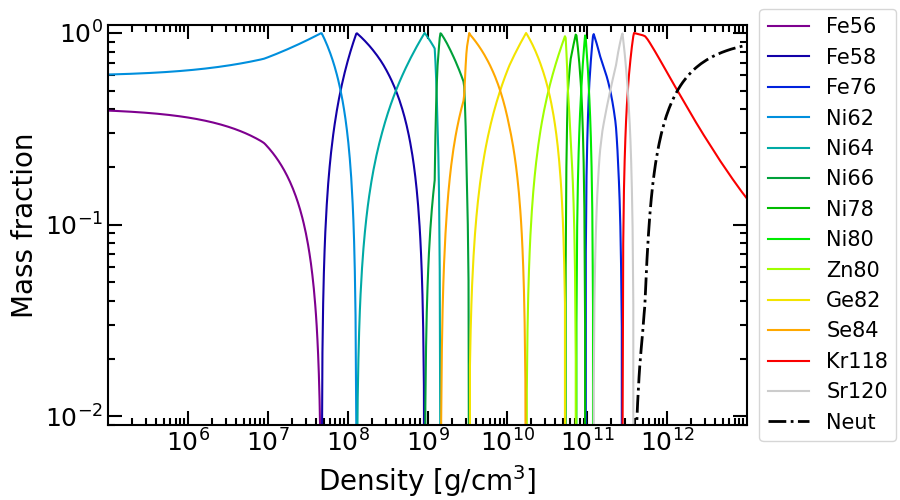}
    \caption{
    Initial mass fractions of isotopes at NSE for the neutron star crust-like matter. 
    Ni62 and Fe56 are the major isotopes near the surface of the neutron star. 
    The electron fraction decreases gradually as density increases, and neutron-rich isotopes appear correspondingly.
    Free neutrons are dripped when the density is above $\sim4\times10^{11}$\gcc\, and become the predominant composition near the threshold density $\rho_{\text{th}}\equiv10^{13}$\gcc.
    }
    \label{fig:initial NSE solution}
\end{figure}

\subsection{Composition Evolution}
\label{sec:Composional evolution}

The chemical composition evolution of neutron star matter is computed individually for each mass element. 
The Maxwell-Boltzmann equation is solved under the given temperature, density, and electron fraction to assign the mass fraction of each isotope in the nuclear reaction network as the initial composition at NSE, which is shown in Figure~\ref{fig:initial NSE solution}. 
A threshold density $\rho_{\text{th}}\equiv10^{13}$\gcc\, is defined. 
A mass element is regarded as core-like (crust-like) matter if its density is above (below) $\rho_{\text{th}}$. 
Above $\rho_{\text{th}}$, the neutron star matter is composed of predominantly free neutrons, with a mass fraction $\gtrsim0.9$, and nuclei forming exotic nuclear structures, such as pasta-like configurations, are immersed. 
The $\beta$-decay channels of nuclei are all blocked by the high electron chemical potential ($\mu_{e} \gtrsim 30$\,MeV, see Equation~(\ref{eq:beta decay})). 
Therefore, we do not determine the composition of mass elements with a density exceeding $\rho_{\text{th}}$, and we neglect the nuclear reactions involved. 
The network is not activated for the mass elements composed of core-like matter in the hydrodynamic simulations.

\begin{figure}[t!]
    %\plotone{figure4.png}
    \includegraphics[width=\columnwidth]{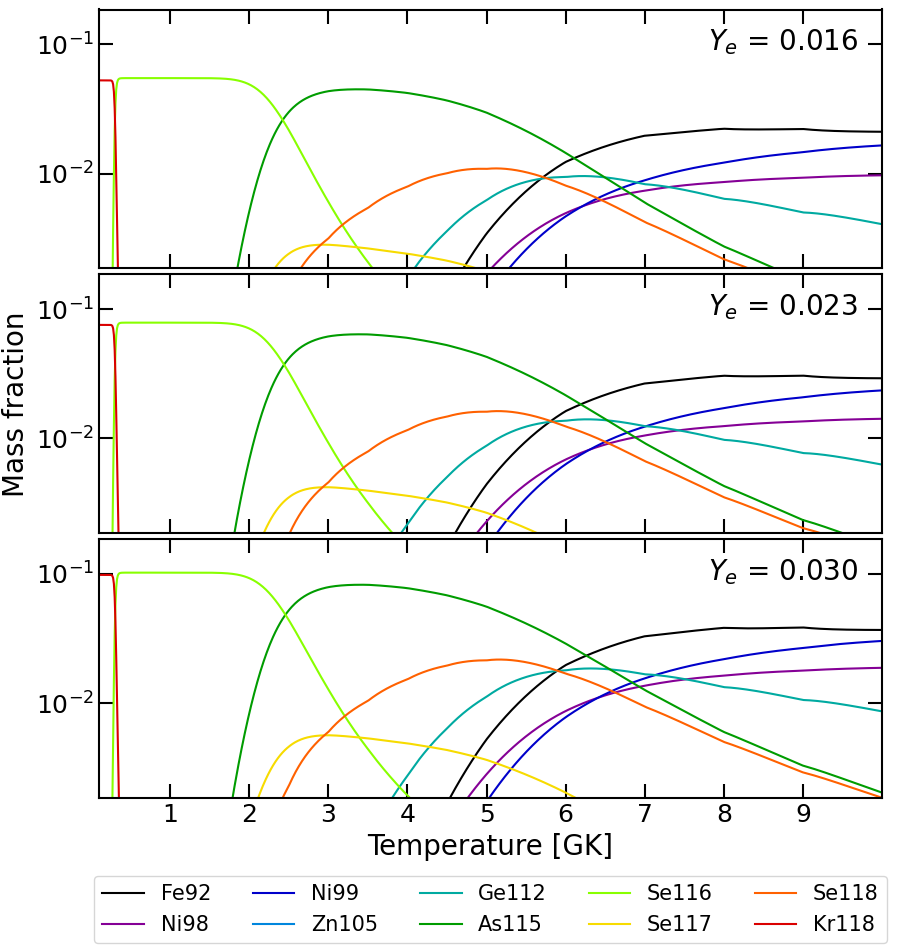}
    \caption{
    Mass fractions of isotopes at NSE and threshold density $\rho_{\text{th}}\equiv10^{13}$\gcc. 
    The solutions at different temperatures and electron fractions $Y_e$ are shown. 
    The mass fraction of the free neutron is $\sim 0.9$\, (not shown in the figure) for the conditions concerned. 
    For $Y_e$ between $0.016$ and $0.030$, the solutions at NSE are alike.
    }
    \label{fig:threshold density NSE solution}
\end{figure}

The maximum temperature reached during the explosion of a subminimal neutron star is not high enough for efficient neutrino capture \citep[][]{PhysRevC.58.3696} or positron capture \citep[][]{RUFFINI20101} to contribute significantly to leptonization of the neutron star matter. 
The modified URCA processes are the major weak interactions between the free neutrons and protons, which are characterized by a timescale much longer than the relevant hydrodynamic timescale \citep[][]{1989ApJ...339..318C}. 
Hence, the $\beta$-decays of nuclei discussed in Section~\ref{sec:nuclear reaction network} are the only weak interactions that alter the electron fractions of the mass elements. 
As a result, the electron fractions of mass elements regarded as core-like matter are fixed to the initial values obtained assuming $\beta$-equilibrium. 
When the density of a mass element drops below $\rho_{\text{th}}$, it transforms from core-like matter to crust-like matter. 
It is assumed that NSE is established after the transition from an exotic nuclear structure, such as pasta-like configurations, to a mixture of nuclei, free neutrons, and electrons as crust-like matter at that time step. 
The mass fractions of isotopes at NSE are assigned to be the composition of the mass element at the temperature, density, and electron fraction after the transition (see Figure~\ref{fig:threshold density NSE solution}). 
After that, the composition of the mass element is obtained by solving the network coupled to the Euler equations.

The full network with all nuclear reactions considered is solved together with the Euler equations in order to investigate their effects consistently. 
It is believed that the leptonization owing to $\beta$-decays of nuclei and an overall net thermal energy generation by nuclear reactions enhance the expansion of the crust-like matter significantly. 
Hence, the leptonization and thermal energy generation are crucially responsible for the onset of hydrodynamical instability. 
The subminimal neutron star becomes unstable and explodes ultimately when it can no longer adjust itself against radial oscillations, and all mass elements are ejected. 
We perform the hydrodynamic simulation until the density of all the mass elements drops near the minimum value available by the EoS table ($\sim 10^{3}$\gcc). 
After that, we extrapolate\footnote{The extrapolation starts when the density of the mass elements reaches $\sim10^{4}$\gcc. To trace the evolution of all mass elements until the extrapolation can be started, we remove the outermost mass element whenever its density reaches near the minimum density available and continue the hydrodynamic simulation of the remaining mass elements up to about $1-1.5$\,s.} the density and temperature of the ejecta assuming homologous and adiabatic expansion described by $\rho(t) \propto t^{-3}$ and $T(t) \propto t^{-1}$ as functions of time $t$. 
We update the network for $1$\,Gyr after the hydrodynamic simulation to determine the stable elements produced by the explosion of the subminimal neutron star.

%----------------------------------------------------------------------
\section{Results}
\label{sec:Results}

We have checked that the initial model of the minimum-mass neutron star constructed is stable in the hydrodynamic simulation coupled to the nuclear reaction network if no perturbation is imposed. 
To initiate an instability, we remove several mass elements from the surface of the star. 
We vary $\Delta S$, the initial mass removal ratio to the total mass of the star, from $0.40$ to $0.30$ in order to mimic the unstable mass transfer in the final approach of the binary system. 
Also, we set the fission fragment asymmetry parameter $\gamma$ to be $0.50$, $0.45$, and $0.40$ to study the effect of asymmetric fission fragment approximation (see Section~\ref{sec:nuclear reaction network}). 
The models are named according to $\Delta S$ and $\gamma$ as L-$\Delta S$-$\gamma$.

\subsection{Explosion of Subminimal Neutron Star}
\label{sec:Explosion of subminimal neutron star}

\begin{figure*}[t!]
    \plotone{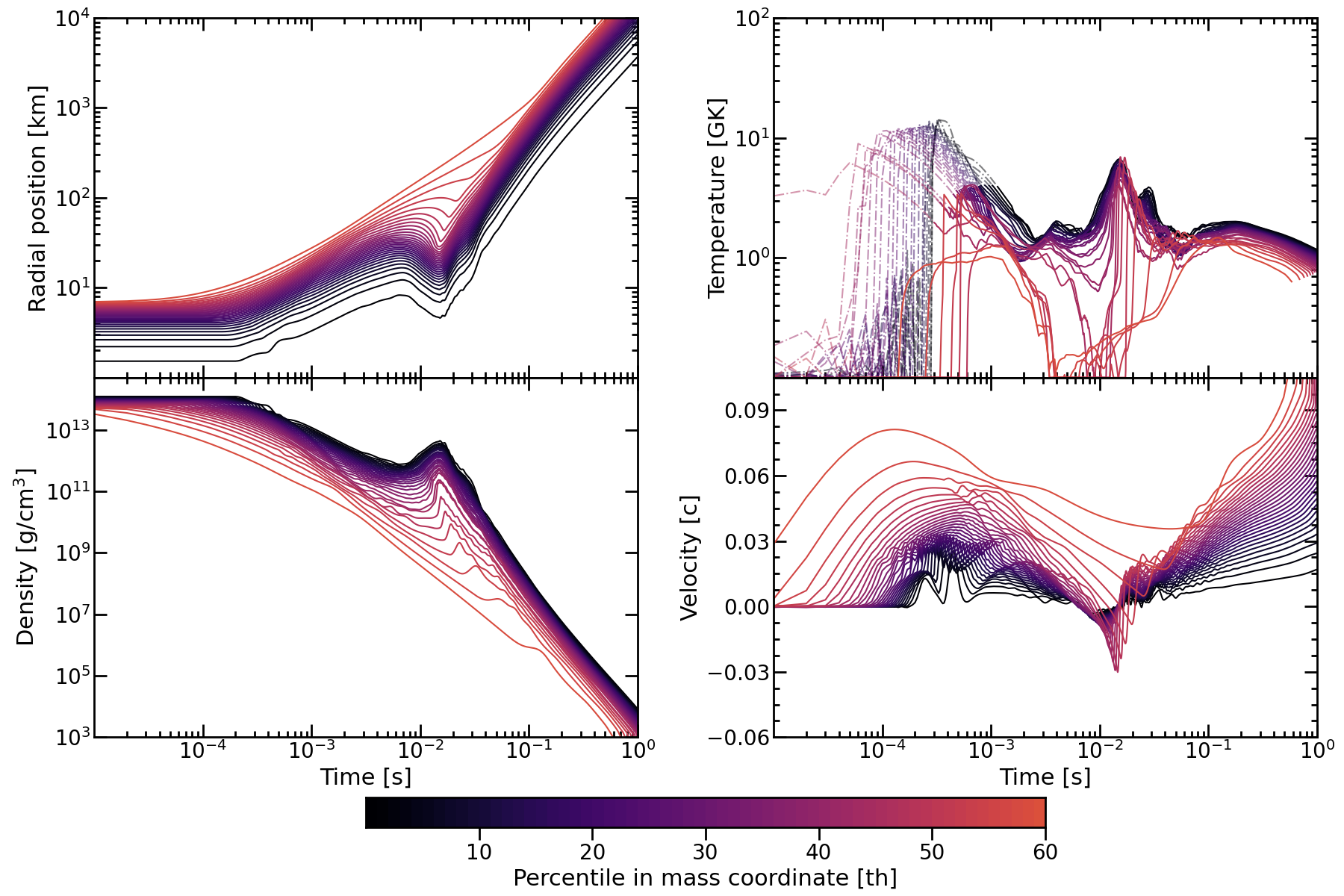}
    \caption{
    Radial positions (upper left panel), densities (lower left panel), temperatures (upper right panel), and velocities (lower right panel) of mass elements vs. time in logarithm scale in the model L-40-50. 
    The temperatures are indicated by dashed-dotted lines prior to the activation of the nuclear reaction network to signal potentially large uncertainty due to the EoS construction ambiguity (see footnote \ref{footnote:phase transition}). 
    After the mass removal, the mass elements near the surface of the exposed neutron star core-like matter start expanding promptly, though most of them remain gravitationally bound. 
    They fall back onto the neutron star at $\sim0.01$\,s and undergo radial oscillation. 
    A new equilibrium of the subminimal neutron star with lower central density is temporarily found between $\sim0.01-0.05$\,s. 
    Nonetheless, the nuclear reaction network is activated as the density of the mass elements is lower than the threshold density $\rho_{\text{th}}\equiv10^{13}$\gcc. 
    The heating effect and leptonization caused by the nuclear reactions alter the structure of the neutron star in a quasi-equilibrium state. 
    The star loses stability after $\sim0.05$\,s, leading to a delayed explosion.
    }
    \label{fig:hydrodynamic L-40-50}
\end{figure*}

The hydrodynamic simulation results of the model L-40-50 are displayed in Figure~\ref{fig:hydrodynamic L-40-50}. 
By imposing the initial mass removal, all the neutron star crust-like matter and part of the neutron star core-like matter are removed. 
The pressure gradient on the surface of the exposed core-like matter increases suddenly, and hence hydrostatic equilibrium is no longer maintained. 
The mass elements expand sequentially from the surface to the center. 
The perturbation made here, however, does not lead to a direct explosion of the whole configuration. 
Most of the mass elements oscillate and settle down at a larger radial position and a lower density temporarily after $\sim0.01$\,s. 
The subminimal neutron star remains at quasi-equilibrium. 
If no nuclear reaction is included, the star oscillates around the new equilibrium position with most of the mass elements remaining gravitationally bounded.

\begin{figure}[t!]
    %\plotone{figure6.png}
    \includegraphics[width=\columnwidth]{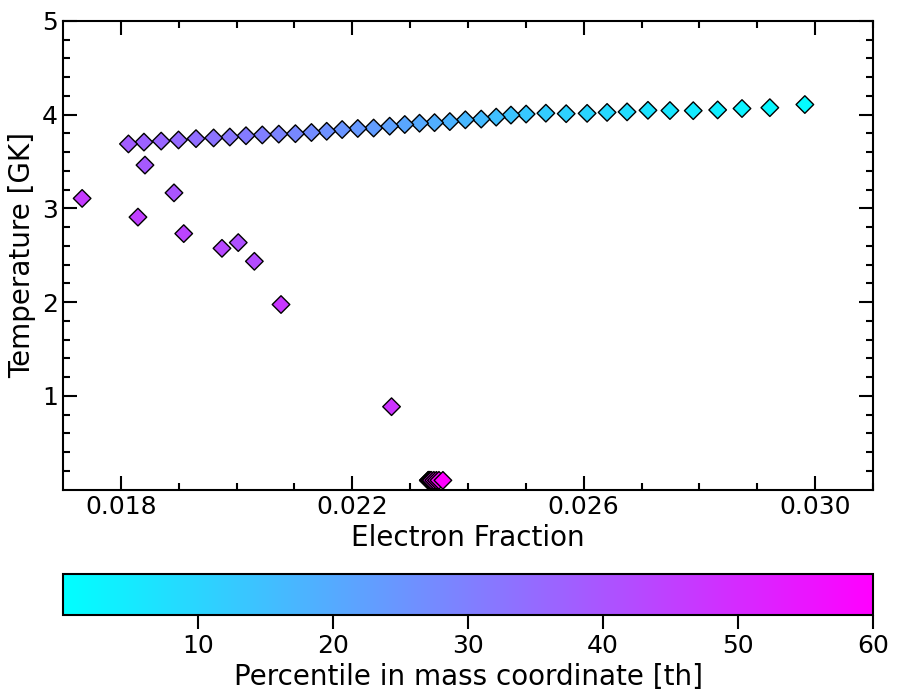}
    \caption{
    Temperatures and electron fractions of the mass elements during the transition from core-like matter to crust-like matter in the model L-40-50. 
    When the density of mass elements drops from $\sim10^{14}$\gcc\, to $\rho_{\text{th}}\equiv10^{13}$\gcc\, due to expansion, a transition from core-like matter to crust-like matter occurs.
    }
    \label{fig:transition temperature and electron fraction}
\end{figure}

\begin{figure}[t!]
    %\plotone{figure7.png}
    \includegraphics[width=\columnwidth]{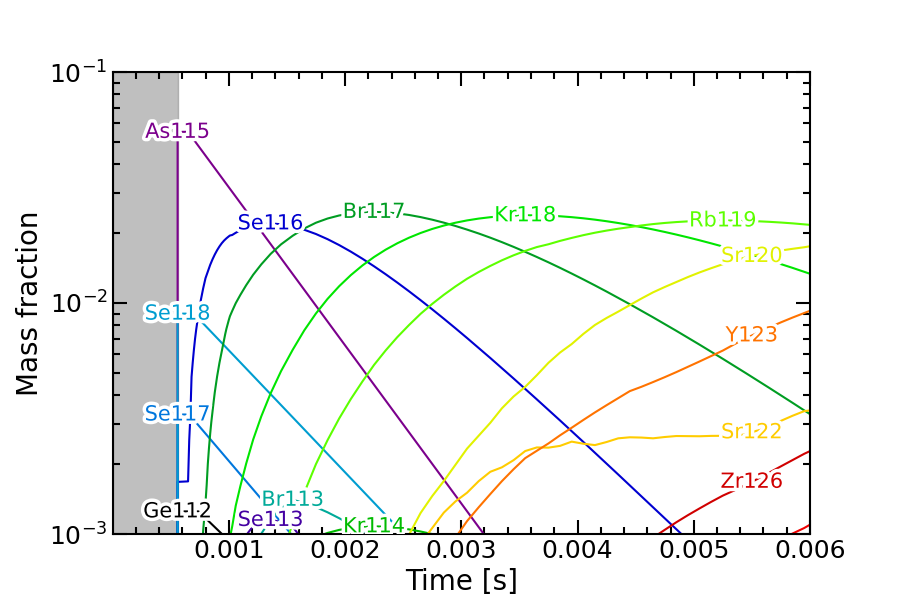}
    \caption{
    Evolution of the isotope abundances of the mass element in the 30th percentile in mass coordinate vs. time in the model L-40-50. 
    After the transition at $\sim0.0006$\,s, As115 is the major nucleus present at NSE. 
    The nuclear reactions, predominately neutron capture and $\beta$-decay, promote the subsequent formations of Se116, Br117, Kr118, and heavier isotopes within a short timescale. 
    The mass fraction of free neutron drops only by $\sim0.5\%$ (not shown in the figure) within the duration shown. 
    For other mass elements with lower temperature during the transition, either Se116 or Kr118 is the major nucleus instead. 
    These two isotopes serve as seed nuclei that undergo $r$-process with very similar evolutionary tracks as the one shown for As115. 
    As a result, whether Kr118, Se116, or As115 is the major nucleus at NSE does not significantly affect the $r$-process later on.
    }
    \label{fig:NSE does not matter}
\end{figure}

Meanwhile, the nuclear reaction network is activated, as the densities of these mass elements are now lower than $\rho_{\text{th}}$. 
The temperatures and electron fractions of the mass elements during the transition from core-like matter to crust-like matter are plotted in Figure~\ref{fig:transition temperature and electron fraction}. 
As discussed in Section~\ref{sec:Composional evolution}, the NSE isotope abundances right after the transition are assigned. 
Note that the temperatures of some mass elements are much lower than $2\times10^9$\,K, the typical minimum temperature that the NSE approximation is regarded as valid during the transition. 
We assume that the NSE approximation still holds for these mass elements. 
Figure~\ref{fig:NSE does not matter} illustrates the evolution of the isotope abundance of a mass element in the 30th percentile in mass coordinates after the transition as an example. 
The differences among the assigned compositions of the mass elements at different temperatures are eliminated by a robust $r$-process very soon. 
Therefore, the simulation results are not noticeably affected even if the NSE approximation may not be accurate at low temperatures.

The $r$-process proceeds efficiently in all mass elements with evolutionary tracks analogous to the one displayed in Figure~\ref{fig:NSE does not matter}. 
The nuclear reactions increase the internal energy and electron fractions of these mass elements, which provide higher thermal pressure and electron degeneracy pressure that alter the hydrodynamic properties of the neutron star. 
The cumulative effects promote the expansion of mass elements, and the subminimal neutron star loses stability gradually. 
A delayed explosion of the whole configuration begins at $\sim0.05$\,s, with all mass elements becoming unbounded as ejecta. 
The radial velocity of the ejecta increases monotonically after the explosion due to the thermal energy generation and leptonization associated with the ongoing nuclear reactions. 
The total kinetic energy and the magnitude of the gravitational potential of the ejecta are $\sim1.4\times10^{50}$\,erg and $\sim6\times10^{47}$\,erg, respectively, before the density of the outermost mass element drops below the minimum density allowed by the EoS table (at $\sim0.5$\,s after the mass removal).

When the subminimal neutron star expands, a rapid rise in temperature up to $\sim10^{10}$\,K\footnote{For the EoS we adopt, there is a first-order phase transition from uniform to nonuniform nuclear matter configurations at density near $6 - 8\times10^{13}$\gcc. During the EoS construction, the phase with lower free energy is chosen if there are solutions for both uniform and nonuniform phases. In some rare cases, the phase with slightly higher free energy is chosen to avoid an unphysical adiabatic index. Hence, the magnitude of temperature rise in our simulations is uncertain because the EoS near the phase transition point is rather ambiguous at low temperatures. For more details, please refer to the original literature on the EoS construction \citep[][]{PhysRevC.96.065802}.\label{footnote:phase transition}}, while the network is not yet activated, is observed at densities above $\rho_{\text{th}}$ (see Figure~\ref{fig:hydrodynamic L-40-50}).
Correspondingly, the mass elements reach $\rho_{\text{th}}$ at different temperatures as illustrated in Figure~\ref{fig:transition temperature and electron fraction}. 
The thermodynamic properties of neutron star matter, however, are insensitive to temperature below $\sim10^{10}$\,K in our simulations\footnote{We have verified that the pressure and the specific internal energy density are insensitive to temperatures below $\sim10^{10}$\,K at densities above $\rho_{\text{th}}$, so the hydrodynamics is also unaffected.}. 
Since the temperature rise may be due to ambiguities in the EoS (see footnote \ref{footnote:phase transition}) and has no effect on the hydrodynamics, we only focus on the nuclear reactions involved below $\rho_{\text{th}}$. 
After the network is activated, the mass elements are powered by the nuclear reactions. 
The high temperature of the crust-like matter, up to about $6\times10^{9}$\,K, reached before the neutron star explodes is attributed to not only the heating effect by the nuclear reactions, but also the adiabatic compression during radial oscillations.

The hydrodynamic evolution of the subminimal neutron star is insensitive to the values of $\gamma$ chosen in the models presented. 
The explosion timescale, however, varies significantly among the models with different values of $\Delta S$. 
With a lower $\Delta S$, the explosion is postponed and the pulsating phase extends substantially. 
It takes a long time for the nuclear reactions to initiate an instability and lead to the explosion, which is indicated by the rapidly decreasing central density. 
After that, the star explodes in a similar way as illustrated in Figure~\ref{fig:hydrodynamic L-40-50}. 
The hydrodynamic simulation results of the models with smaller values of $\Delta S$ are presented in Appendix~\ref{app: other models} for reference.

\subsection{Neutrino Emission and Electromagnetic Radiation}
\label{sec:Neutrino emission and electromagnetic radiation}

\begin{figure}[t!]
    %\plotone{figure8.png}
    \includegraphics[width=\columnwidth]{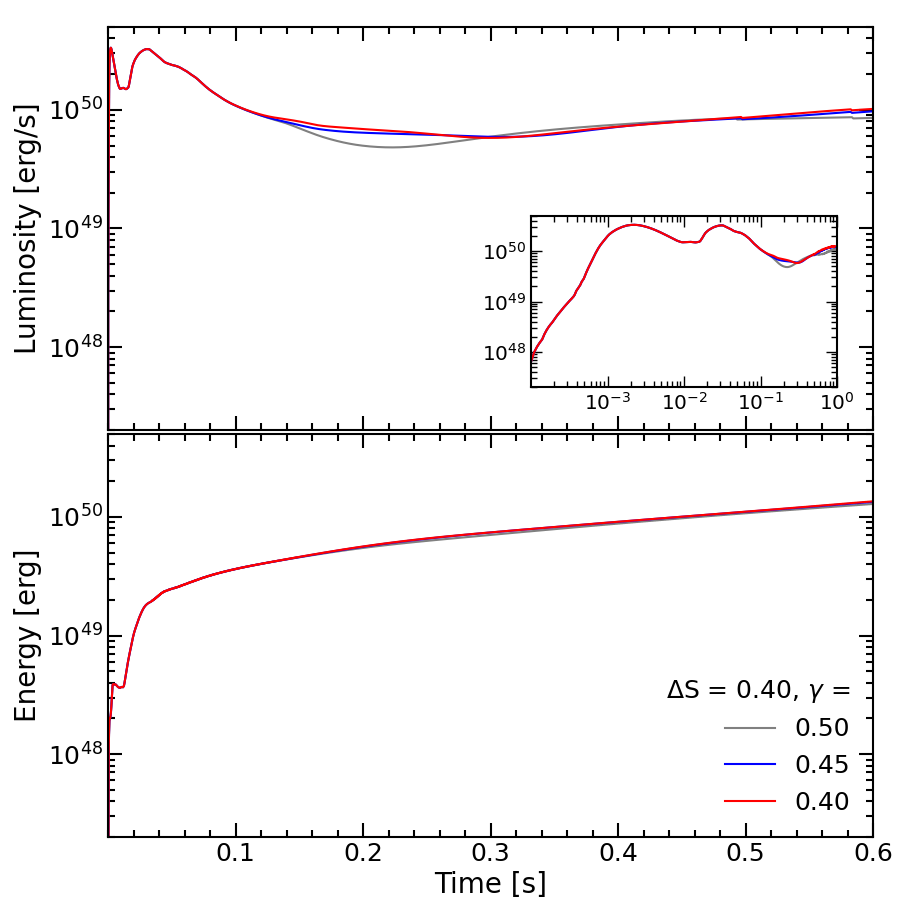}
    \caption{
    Upper panel: electron antineutrino luminosity emitted through $\beta$-decays of nuclei in hydrodynamic simulation vs. time in the models L-40-50, L-40-45, and L-40-40. 
    A subplot zooms in on the initial peak and shows that the peak luminosity is $\sim3\times10^{50}$\ergs\, in the three models. 
    Lower panel: net cumulative thermal energy injected into the neutron star by the nuclear reaction network vs. time in the same set of models.
    }
    \label{fig:network energy among gamma}
\end{figure}

After the explosion of the neutron star, the nuclear reactions, mainly $\beta$-decays of nuclei, power the ejecta continuously. 
The temperature of the ejecta is sustained at $\sim10^9$\,K and decreases slowly because of adiabatic cooling. 
The time evolution of electron antineutrino luminosity and the net cumulative thermal energy deposited on the subminimal neutron star by the network in the models L-40-50, L-40-45, and L-40-40 are shown in Figure~\ref{fig:network energy among gamma}. 
The electron antineutrino luminosity associated with $\beta$-decays of nuclei is predicted using Equation~(\ref{eq:neutrino energy}) during the explosion of a subminimal neutron star. 
A peak luminosity of $\sim3\times10^{50}$\ergs\, is reached shortly after the mass removal in the three models. 
The little trough in the luminosity at $\sim0.01$\,s is caused by the transient rise in electron chemical potential (see Equation~(\ref{eq:beta decay})) associated with adiabatic compression during radial oscillations of the star. 
After the peak in neutrino emission, the average luminosity is maintained at $\sim10^{50}$\ergs. 
Meanwhile, the net cumulative thermal energy deposited on the subminimal neutron star by the network is $\sim10^{50}$\,erg in the three models before the density of the outermost mass element drops below the minimum density provided by the EoS table at $\sim0.5$\,s. 
The hydrodynamic evolution of models L-40-45 and L-40-40 is very similar to that of the model L-40-50. 
Also, the neutrino luminosities of the three models are very similar in general. 
These results are, therefore, insensitive to the variation in fission fragment asymmetry parameter $\gamma$. 
The electron antineutrino luminosity versus time and the net cumulative thermal energy deposited on the neutron star by the network in the models L-40-50, L-38-50, L-36-50, L-34-50, L-32-50, and L-30-50 are also shown in Appendix~\ref{app: numerical test} as a comparison.

\begin{figure}[b!]
    %\plotone{figure9.png}
    \includegraphics[width=\columnwidth]{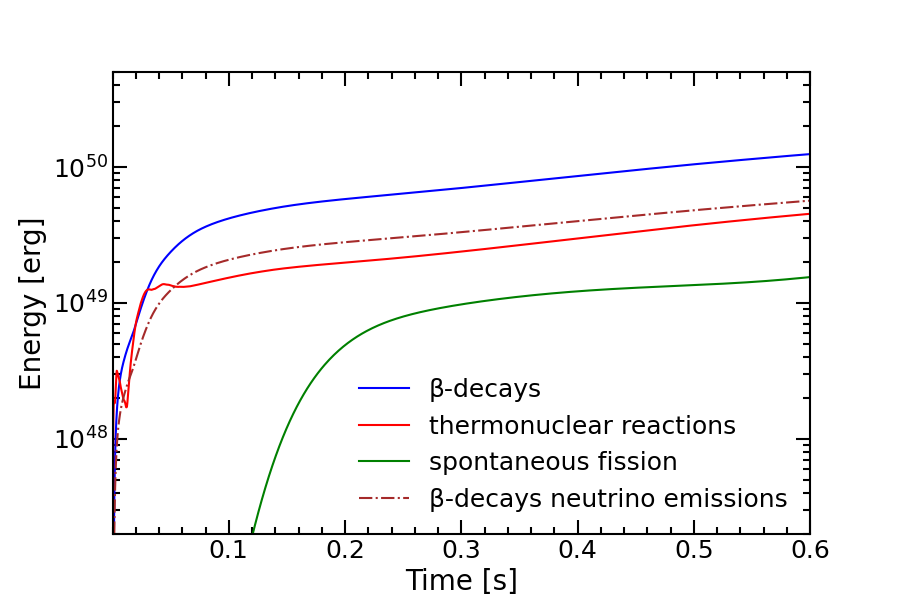}
    \caption{
    Cumulative energy source terms by the nuclear reaction network vs. time in the model L-40-50. 
    The absolute value of the cooling term corresponding to $\beta$-decay neutrino emissions is plotted as a brown dashed-dotted line.
    }
    \label{fig:network energy for L-40-50}
\end{figure}

\begin{figure}[b!]
    %\plotone{figure10.png}
    \includegraphics[width=\columnwidth]{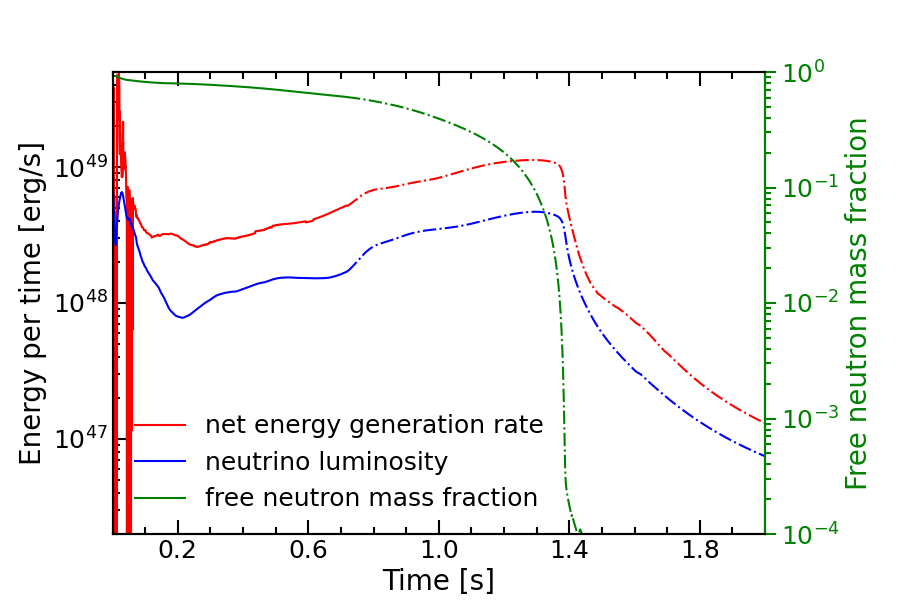}
    \caption{
    Net thermal energy generation rate, local electron antineutrino luminosity emitted through $\beta$-decays of nuclei, and the free neutron mass fraction of the mass element in the 30th percentile in mass coordinate vs. time in the model L-40-50. 
    The solid curves are obtained directly from hydrodynamic simulations, and the dashed curves are obtained by extrapolation of the density and temperature.
    }
    \label{fig:extrapolated energy per time}
\end{figure}

In the model L-40-50, we evaluate the cumulative energy source terms contributed by $\beta$-decays, thermonuclear reactions, spontaneous fission, and neutrino emissions through $\beta$-decay of nuclei versus time (see Figure~\ref{fig:network energy for L-40-50}). 
It is found that $\beta$-decay is the dominating thermal energy source throughout the simulation, while the thermonuclear reactions and spontaneous fission of nuclei are rather insignificant. 
From the nucleosynthesis calculations, we notice that the net thermal energy generation rate and neutrino luminosity are sustained until $\sim1.4$\,s and plunge when the mass fraction of free neutron vanishes (see Figure~\ref{fig:extrapolated energy per time} for the evolution of relevant quantities of the mass element in the 30th percentile in mass coordinate as an example). 
The temperature of the ejecta remains high, and the expansion is continuously powered by the nuclear reactions until the free neutrons are exhausted, and the production of neutron-rich nuclei with short $\beta$-decay half-lives is terminated. 
An electromagnetic peak near soft gamma-ray lasting for a few seconds in total is expected accordingly, assuming blackbody emission. 
Likewise, the neutrino emission should last for a few seconds and diminish rapidly following the termination of $r$-process.

\subsection{Nucleosynthesis}
\label{sec:Nucleosynthesis}

During the explosion of a subminimal neutron star, the neutron star matter in the core, with initially very low electron fraction, is decompressed without experiencing very strong heating that can be achieved in other astronomical $r$-process sites, such as core-collapse supernovae and binary neutron star mergers, where the temperature can reach the order of $10^{11}$\,K. 
The maximum temperature experienced by the mass elements in the simulation is $\sim10^{10}$\,K, which is much lower than the typical temperature scale in the former two scenarios. 
A robust $r$-process occurs because of the initially high neutron excess and low electron fraction of the ejecta. 
Beyond the simulation time, we extend the nucleosynthesis calculations by extrapolating the density and temperature of the ejecta, assuming free expansion. 
It is noticed that all mass elements experience similar evolutionary tracks of composition in the models L-40-50, L-40-45, and L-40-40. 
The temperatures and electron fractions of the mass elements in the three models when the network starts operating are listed in Table~\ref{tab:mass element phase transition}. 
The inner ejecta, originating from the neutron star core region, has a relatively high temperature and electron fraction compared to the other two categories after the decompression. 
The intermediate ejecta experiences moderate heating and has a low initial electron fraction. 
The outer ejecta, with almost no adiabatic compression during the radial oscillations, is initially cold and has moderate electron fraction.

\begin{figure*}[t!] 
    \centering
    \includegraphics[width=0.82\textwidth]{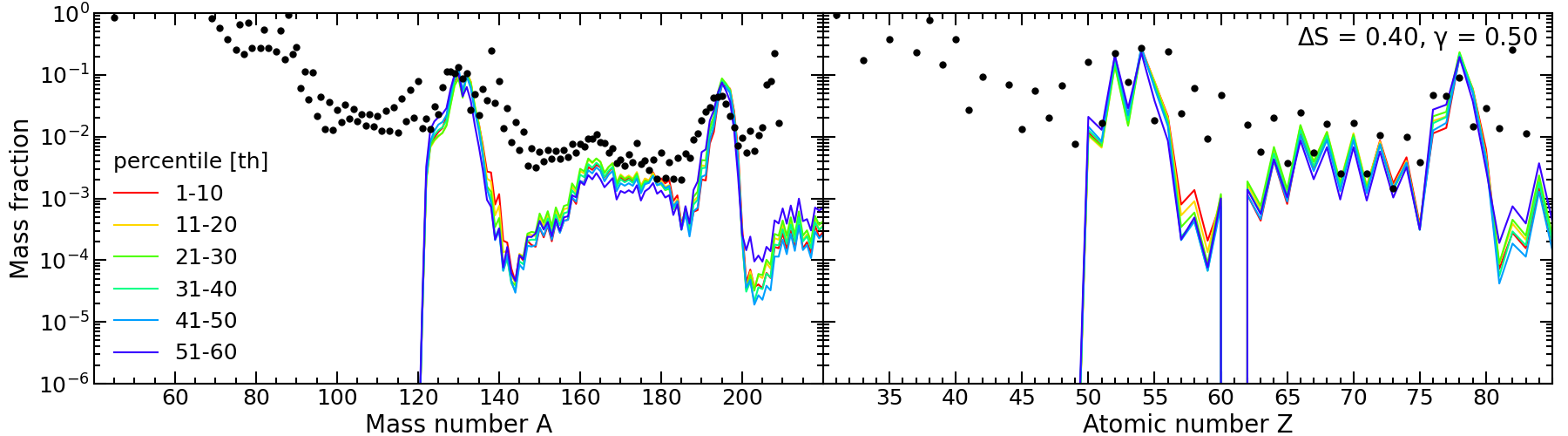}
    \caption{
    Averaged fractions of isotopes vs. mass number $A$ and atomic number $Z$ of every 10 mass elements at the end of nucleosynthesis calculation in the model L-40-50. 
    The solar abundance distribution \citep[][]{lodders2019solar}, scaled to match with the averaged mass fraction of Xe132 of the mass elements from the 1st to 10th percentile in mass coordinate, is shown by the black scatters.
    }
    \label{fig:nucleosynthesis L-40-50}
\end{figure*}

\begin{figure*}[t!]
    \centering
    \includegraphics[width=0.82\textwidth]{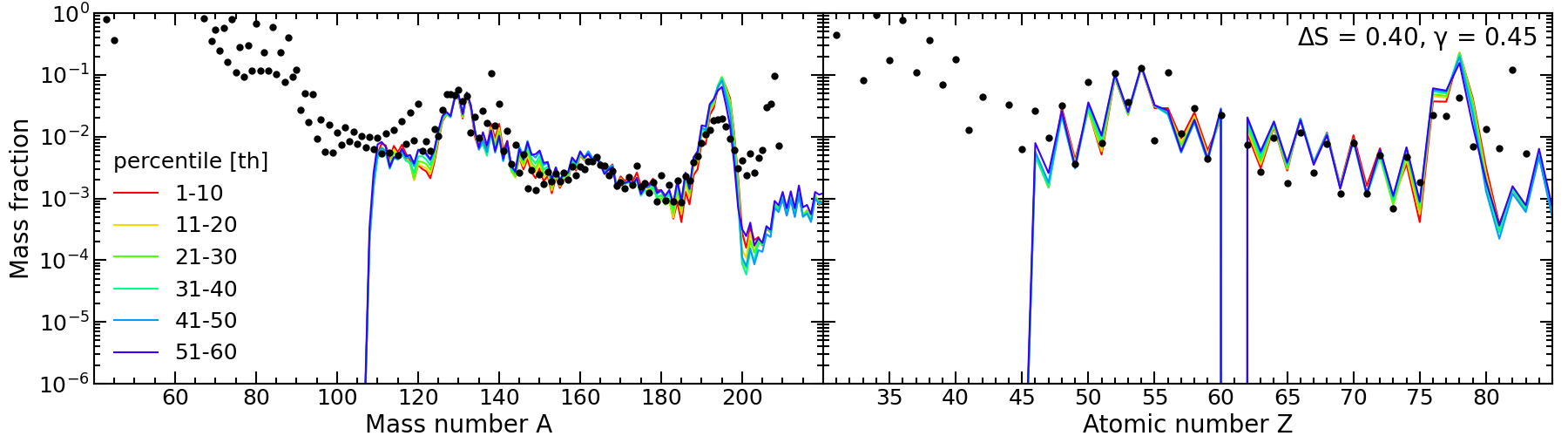}
    \caption{Same as Figure~\ref{fig:nucleosynthesis L-40-50}, but for the model L-40-45.}
    \label{fig:nucleosynthesis L-40-45}
\end{figure*}

\begin{figure*}[t!]
    \centering
    \includegraphics[width=0.82\textwidth]{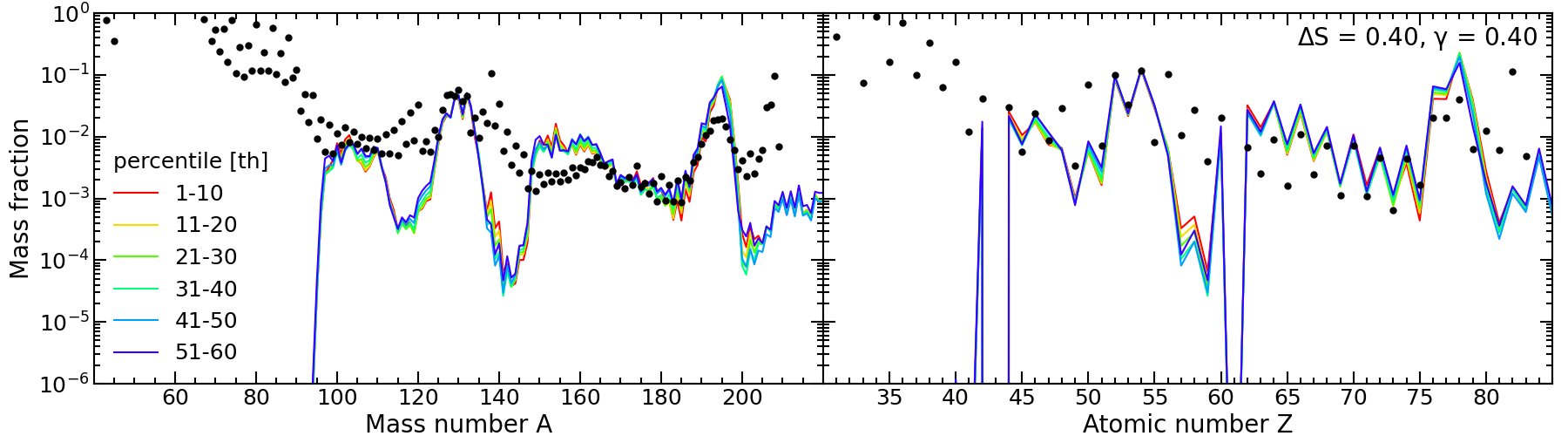}
    \caption{Same as Figure~\ref{fig:nucleosynthesis L-40-50}, but for the model L-40-40.}
    \label{fig:nucleosynthesis L-40-40}
\end{figure*}

The mass fractions of isotopes produced in the models L-40-50, L-40-45, and L-40-40 after $1$\,Gyr are plotted versus mass number and atomic number in Figures~\ref{fig:nucleosynthesis L-40-50}-\ref{fig:nucleosynthesis L-40-40}, respectively. 
The solar abundance distribution by \citet{lodders2019solar}, scaled to match with the averaged production peak of Xe132 among the first to tenth mass elements, is shown by the black scatters in the figures for comparison. 
With the occurrence of a robust and long-lasting $r$-process in all mass elements, the composition evolution becomes insensitive to the tiny difference among the initial conditions of these mass elements. 
The end products of nucleosynthesis in all mass elements from the same model are very similar.

On the other hand, the production curves of the three models are distinct because of the different fission fragment asymmetries. 
In the model L-40-50, abundant production of elements near the third $r$-process peak ($A\approx195$), comparable to the solar abundance observation, is found. 
Lanthanides ($Z=57-71$) are insufficiently synthesized, especially for those before gadolinium ($Z=64$). 
Due to the symmetric fission fragment approximation applied, elements with $Z<45$ are not produced after the spontaneous fission of heavy nuclei. 
The end products of nucleosynthesis are lacking in elements before the second $r$-process peak ($A\approx130$). 
In the model L-40-45, the solar abundance distribution is well recovered from the second $r$-process peak to the third $r$-process peak. 
Moreover, an excessive production of elements at the third $r$-process peak is obtained. 
As for the production curve of the model L-40-40, elements slightly less massive than the second $r$-process peak are present. 
However, elements on the two sides of the second $r$-process peak are significantly underproduced when compared to the solar abundance distribution. 
No elements around the first r process peak ($A\approx80$) are formed in the three models.

\begin{figure*}[t!]
    \centering
    \includegraphics[width=0.82\textwidth]{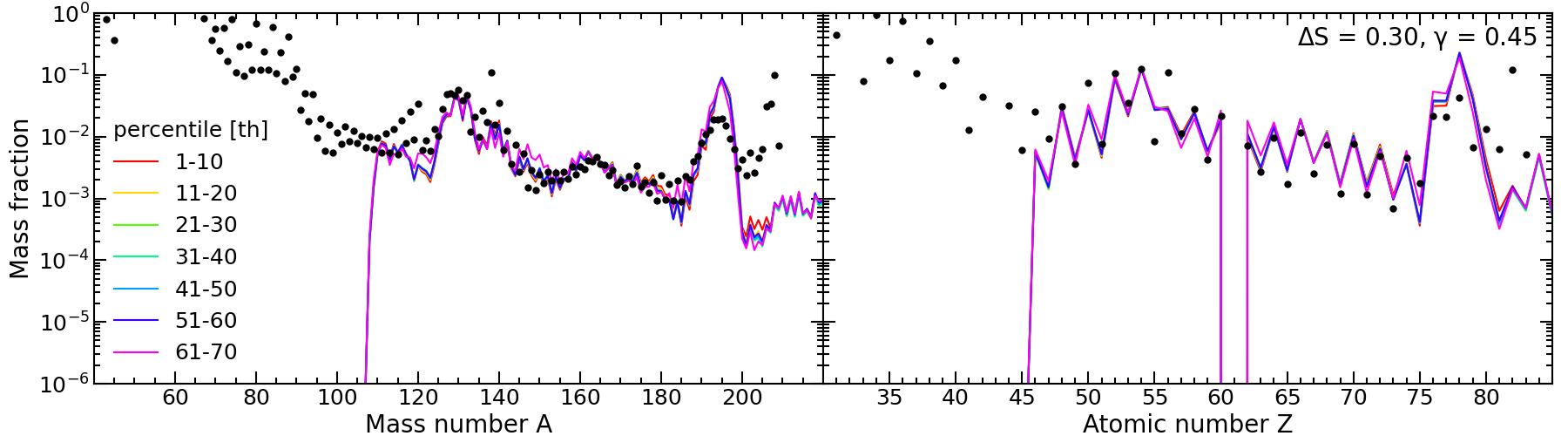}
    \caption{
    Same as Figure~\ref{fig:nucleosynthesis L-40-45}, but for the model L-30-45. 
    The production curves are very similar to those shown in Figure~\ref{fig:nucleosynthesis L-40-45} in general.
    }
    \label{fig:nucleosynthesis L-30-45}
\end{figure*}

Furthermore, the production curves of the model L-30-45 are shown in Figure~\ref{fig:nucleosynthesis L-30-45} to illustrate the robustness of the results. 
The general features of the production curves in the models L-30-45 and L-40-45 are very similar. 
The nucleosynthesis is thus generic against the variation in $\Delta S$ among different models.

\begin{deluxetable}{ccccc}
    \label{tab:mass element phase transition}
    \caption{Nuclear Reaction Network Properties}
    \tablewidth{0pt}
    \tablehead{
    \colhead{Percentile [th]} & \colhead{$Y_{e,\text{ini}}$} & \colhead{$T_{\text{ini}}$ [GK]}  & \colhead{$T_{\text{max}}$ [GK]} &\colhead{Ejecta Category}}
    \startdata
     1-10 & 0.0278 & 4.051 & 6.310 & inner \\
    11-20 & 0.0244 & 3.966 & 5.886 & inner \\
    21-30 & 0.0217 & 3.834 & 5.339 & intermediate \\
    31-40 & 0.0191 & 3.651 & 4.903 & intermediate \\
    41-50 & 0.0205 & 1.946 & 5.6 & intermediate \\
    51-60 & 0.0234 & 0.100 & 3.079 & outer \\
    \enddata
    \tablecomments{
    Percentile in mass coordinate, averaged initial electron fraction $Y_{e,\text{ini}}$, and averaged initial temperature $T_{\text{ini}}$ of the mass elements when the nuclear reaction network starts operation at the threshold density $\rho_{\text{th}}\equiv10^{13}$\gcc\, in the models L-40-50, L-40-45, and L-40-40. 
    The averaged maximum temperature $T_{\text{max}}$ reached by the mass elements after the activation of the network is listed as well. 
    In most of the mass elements, the peak temperature is reached when the first adiabatic compression wave arrives at $\sim0.01$\,s, while nuclear fission takes place only after $\sim0.03$\,s. 
    Therefore, the maximum temperature reached by the mass elements in the three models is nearly identical.
    }
\end{deluxetable}

%----------------------------------------------------------------------
\section{Conclusions}
\label{sec:Conclusions}

We present a study of the explosion of subminimal neutron stars in close neutron star binaries. 
Based on the analysis of mass transfer stability and timescale in asymmetric neutron star binaries \citep[][]{1977ApJ...215..311C,1984SvAL...10..177B}, we construct the initial subminimal neutron star model with the minimum mass allowed by the EoS at equilibrium. 
The hydrodynamic instability is initiated by removing a portion of mass from its surface, following the methodology adopted by \citet{1993ApJ...414..717C} and \citet{1998A&A...334..159S}. 
A modern EoS and a detailed nuclear reaction network are coupled with the Euler equations to study the evolution of the neutron star. 
The major factor responsible for the instability development is leptonization by the $\beta$-decays of nuclei, which results in a higher electron degeneracy pressure of the neutron star matter. 
The heating effect following the $\beta$-decays serves as a supplementary factor that further enhances the expansion of neutron star crust-like matter. 
The overall thermal energy generated through thermonuclear reactions and spontaneous fission is rather minor in comparison. 
The net cumulative thermal energy injected into the neutron star matter from the network is in the order of $10^{50}$\,erg, which is only $\sim0.1-1$\% of its rest mass. 
While nuclear energy is a negligible contribution of energy in other astronomical simulations of typical neutron stars, it plays a significant role in a subminimal neutron star, which is just marginally stable.

The peak luminosity of electron antineutrino emission due to $\beta$-decays of nuclei predicted by our models is $1-2$ order lower than the values reported by \citet{1993ApJ...414..717C} and \citet{1998A&A...334..159S}. 
A maximized choice of the coefficient of $\beta$-decay rate estimation formula with uncertainty \citep[][]{1977ApJ...213..225L} used in their studies is responsible for the difference, and \citet{1998A&A...334..159S} have already pointed out that the $\beta$-decay rate is probably overestimated in their simulations. 
Adopting realistic $\beta$-decay half-lives of nuclei, our calculations predict a neutrino burst with a lower peak in luminosity but lasting for a longer duration. 
The ejecta is powered by the nuclear reactions, and thus the temperature stays at $\sim10^9$\,K despite adiabatic cooling. 
At the end of the simulation, the radius of the outermost mass element grows to $\sim10^4$\,km. 
Radiative cooling may become effective when the surface area of the expanding ejecta is large, and the temperature should decline more quickly. 
Further investigation is required to predict whether a soft gamma-ray burst, with high enough luminosity to be observed from the earth, is associated with the explosion of the subminimal neutron star. 
A bolometric light curve powered by the nuclear reactions may be obtained for further research.

The explosion of subminimal neutron star releases $\sim0.05$\ms\, of ejecta in total. 
The $r$-process occurs in all mass elements, according to our calculations. 
The excessive production of elements near the third $r$-process peak is observed, especially in the models L-40-45 and L-40-40. 
We attribute the synthesis of heavy elements of this event to the initially low electron fraction and the relatively low temperature of the ejecta. 
The two conditions allow the nucleosynthesis to operate under a high neutron excess environment for a sufficiently long duration. 
A similar scenario is also realized in neutron star mergers with high neutron excess. 
The observation of lanthanides and elements at the third $r$-process peak associated with a binary neutron star merger by \citet{2017ApJ...848L..27T} suggests the crucial contribution from neutron star ejecta to the heavy elements production in the Universe. 
The neutron star merger as a promising astronomical $r$-process site is under extensive studies \citep[e.g.,][]{1999ApJ...525L.121F,2017PhRvD..96l4005B,2018ApJ...855...99C,2022arXiv220707421K}.

Meanwhile, several recent researches on binary neutron star mergers and black hole-neutron star mergers suggested that cold dynamical ejecta, originating from the outer regions of neutron stars, may abundantly produce heavy elements up to the third $r$-process peak \citep[e.g.,][]{RevModPhys.93.015002}. 
Our simulation results basically agree with the prediction that cold ejecta from neutron stars is favorable for the formation of elements near the third $r$-process peak \citep[][]{10.1111/j.1365-2966.2012.21859.x}. 
Furthermore, the recent investigation on the nucleosynthesis of subminimal neutron star explosion by \citet{Panov_2020} found that heavy elements up to the third $r$-process peak can be excessively formed in the ejecta from the inner crust initially. 
We notice that such a feature is also present in our results regarding ejecta originating from deeper regions of the subminimal neutron star. 
While the nucleosynthesis calculation in their work is performed without coupling to the hydrodynamic, we may realize that the robustness of the $r$-process during the subminimal neutron star explosion is insensitive to the hydrodynamic evolution of the ejecta with sufficiently low initial electron fraction. 
The total dynamical ejecta mass from binary neutron star mergers is typically in the order of $10^{-3}-10^{-2}$\ms\, \citep[][]{2017PhRvD..96l4005B,Radice_2018,2022MNRAS.510L...7K}, less than what we have obtained from subminimal neutron star explosions. 
Therefore, the latter should also contribute to the heavy element abundances if their event rate is comparable to that of binary neutron star mergers.

We demonstrate that the fission fragment calculation can significantly affect the nucleosynthesis results. 
The fragment asymmetry parameter $\gamma$ is introduced to mimic the fission fragment distribution. 
The study by \citet{Hao_2022} implied that the models with $\gamma$ = $0.45$ or $0.40$ should give more reliable results than those with symmetric fission fragments, such as in the model L-40-50, as well as previous studies by \citet{1993ApJ...414..717C} and \citet{1998A&A...334..159S}. 
Further study may be conducted by determining the fission fragment distribution for each isotope individually with more sophisticated methods \citep[][]{2020PhRvC.101e4607M,2020ApJ...896...28V,Hao_2022}, which should improve the accuracy of the predicted heavy elements production. 
Moreover, neutron-induced \citep[][]{2010A&A...513A..61P} and $\beta$-delayed fission \citep[][]{Panov_2016} may be included. 
Nevertheless, the fission properties, as well as other nuclear reactions involving neutron-rich and heavy nuclei, are not well determined experimentally. 
Theoretical modeling must be used in $r$-process simulation studies, and thus systematic uncertainties always remain in these nucleosynthesis calculations, including the results presented in this paper.

A crucial assumption of our study is the existence of peculiarly light neutron stars with mass $\sim0.1$\ms\, in close neutron star binaries. 
Even though such a configuration is allowed by modern EoS, a neutron star with such a small mass has never been observed \citep[][]{10.1093/mnras/sty2460}. 
Nevertheless, recent research on asymmetric neutron star binaries \citep[e.g.,][]{Ferdman_2020,PhysRevD.101.044053} opens the door to the dynamical formation of low-mass neutron stars through mass transfer. 
In our study, a large fraction of mass ($\Delta S \geq 0.30$) is removed from the surface of the minimum-mass neutron star initially. 
Whether this treatment is realistic in describing the tidal stripping of the low-mass neutron star is unclear. 
We find that the delayed explosion of the subminimal neutron star is considerably postponed if the initial mass removal ratio $\Delta S$ is lower than $0.40$. 
Although the hydrodynamic evolution of the star after losing stability is rather insensitive to the exact value of $\Delta S$, it experiences more periods of radial oscillation during the quasi-equilibrium state if $\Delta S$ is smaller. 
In addition, the temperature of the neutron star crust-like matter temporarily exceeds $10^{10}$\,K whenever the compression wave arrives. 
Neutrino capture and positron capture may additionally give rise to leptonization alongside $\beta$-decays of nuclei. 
Whether the electron fraction remains as low as the value found in our study has to be justified. 
On the other hand, a small value of $\Delta S$ may contradict the conditions described by \citet[][]{1977ApJ...215..311C} and \citet[][]{1984SvAL...10..177B} that the tidal disruption of the whole subminimal neutron star happens at most within seconds. 
While the massive companion is assumed to be stable during the mass transfer in this paper, the influence of the massive companion must be considered if the instability of the subminimal neutron star takes a long duration to develop.

A robust $r$-process nucleosynthesis is realized in the ejecta from the explosion of a subminimal neutron star. 
Lanthanides and heavy elements near the second and third $r$-process peaks are synthesized as end products of the nucleosynthesis, with similar mass fractions as the solar abundances.

%----------------------------------------------------------------------
\section{Acknowledgement}
\label{sec:Acknowledgement}

We acknowledge F. X. Timmes for making the nuclear reaction subroutine Torch open-source and the explanation about its usage. 
We thank A. S. Schneider for making the EoS publicly available. 
We also thank P. Möller and K. P. Santhosh for providing publicly available nuclear reaction rates. 
This work is partially supported by a grant from the Research Grant Council of the Hong Kong Special Administrative Region, China (Project Nos. 14300320 and 14304322). 
This material is based upon work supported by the National Science Foundation under Grant AST-2316807.

\appendix

%----------------------------------------------------------------------
\section{Models with smaller \texorpdfstring{$\Delta S$}{delta S}}
\label{app: other models}

\begin{figure*}[p!]
    \plotone{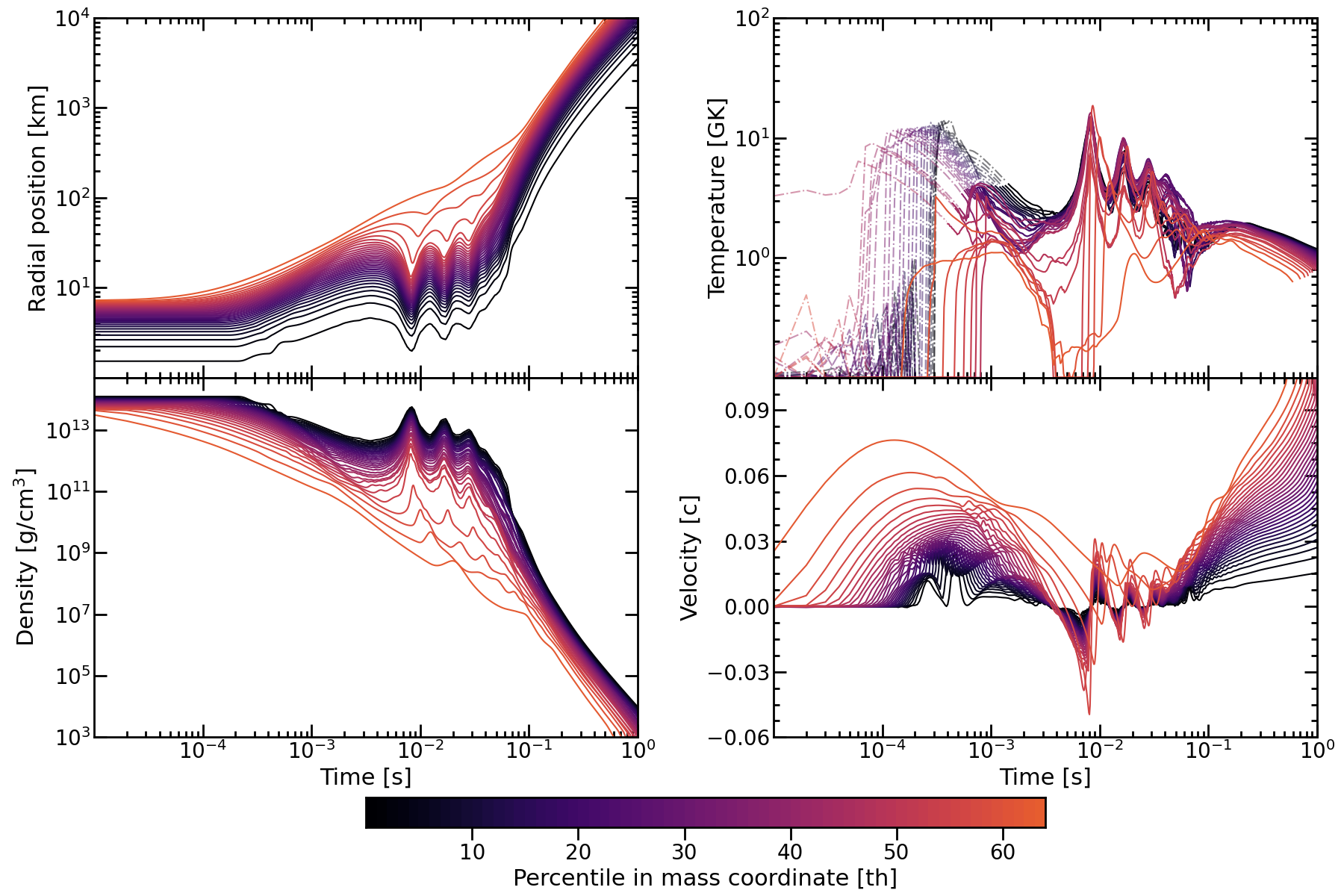}
    \caption{Same as Figure~\ref{fig:hydrodynamic L-40-50}, but for the model L-36-50.}
    \label{fig:hydrodynamic L-36-50}
\end{figure*}

\begin{figure*}
    \plotone{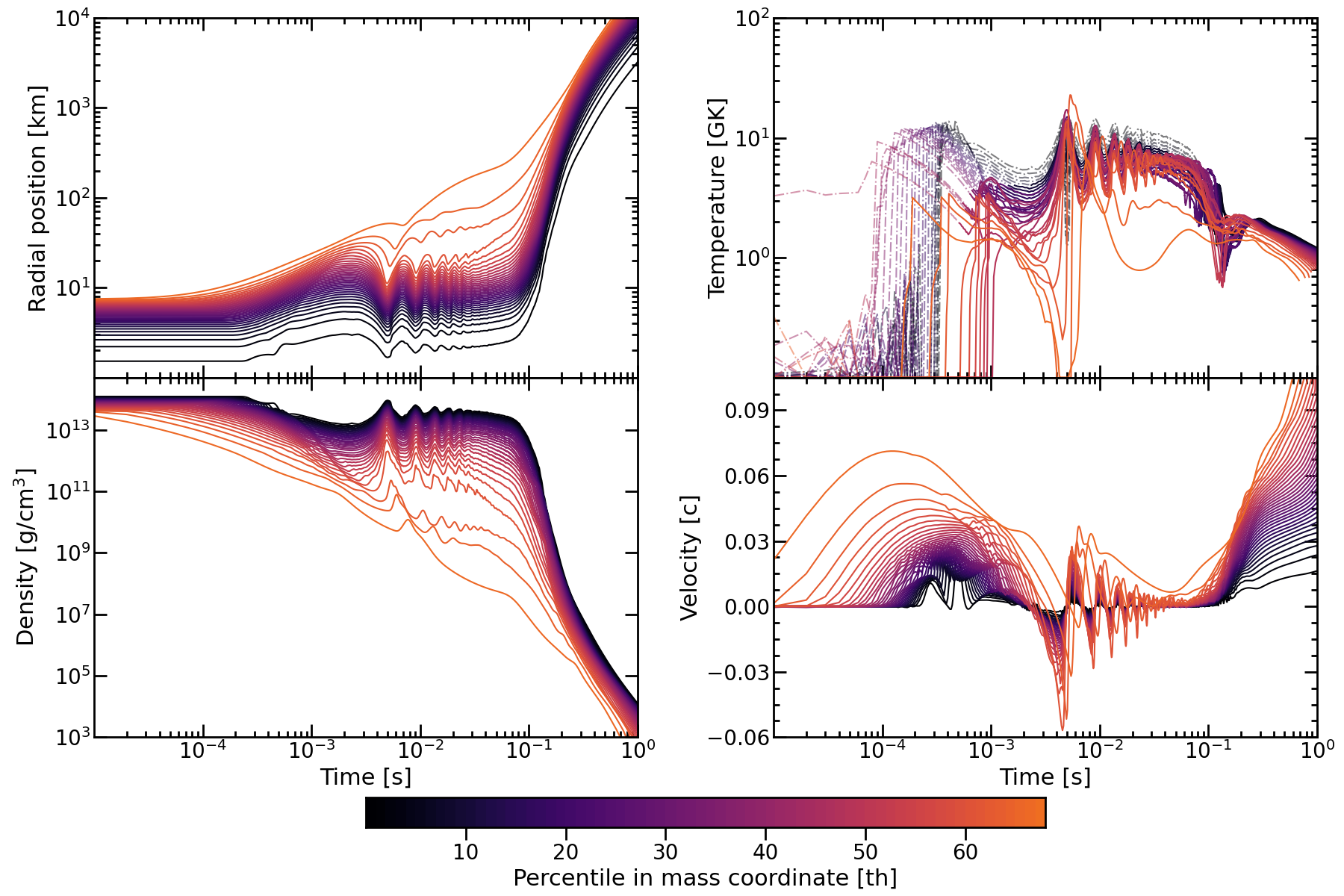}
    \caption{Same as Figure~\ref{fig:hydrodynamic L-40-50}, but for the model L-32-50.}
    \label{fig:hydrodynamic L-32-50}
\end{figure*}

The hydrodynamic evolution of the model L-40-50 is discussed in detail in Section~\ref{sec:Explosion of subminimal neutron star}. 
As mentioned previously, the explosion is further postponed for models with lower $\Delta S$. 
Consequently, the temperature reached by the mass elements can be even higher during radial oscillations prior to the explosion. 
The hydrodynamic simulation results of the models L-36-50 and L-32-50 are shown in Figures~\ref{fig:hydrodynamic L-36-50} and \ref{fig:hydrodynamic L-32-50}, respectively, to illustrate the influence of a smaller $\Delta S$ in the hydrodynamic simulations.

\begin{figure}[t!]
    %\plotone{figure17.png}
    \includegraphics[width=\columnwidth]{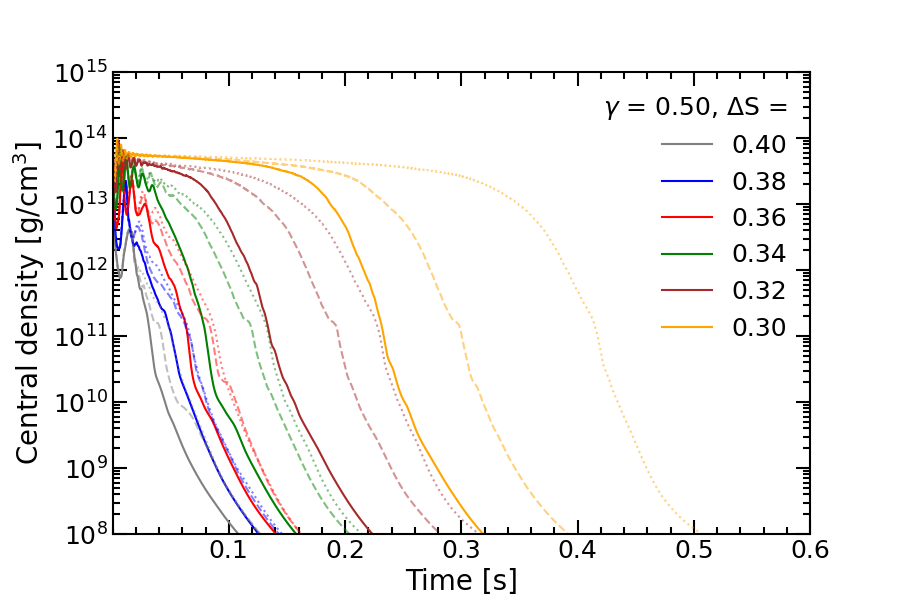}
    \caption{
    Evolution of central densities in the models L-40-50, L-38-50, L-36-50, L-34-50, L-32-50, and L-30-50. 
    The nuclear reaction network is updated for every $\delta t_{\text{net}} =$ $1\times10^{-4}$\,s (solid lines), $2\times10^{-4}$\,s (dashed lines), and $5\times10^{-4}$\,s (dotted lines) before the explosions.
    }
    \label{fig:central density among delta S}
\end{figure}

\begin{figure}[t!]
    %\plotone{figure10.png}
    \includegraphics[width=\columnwidth]{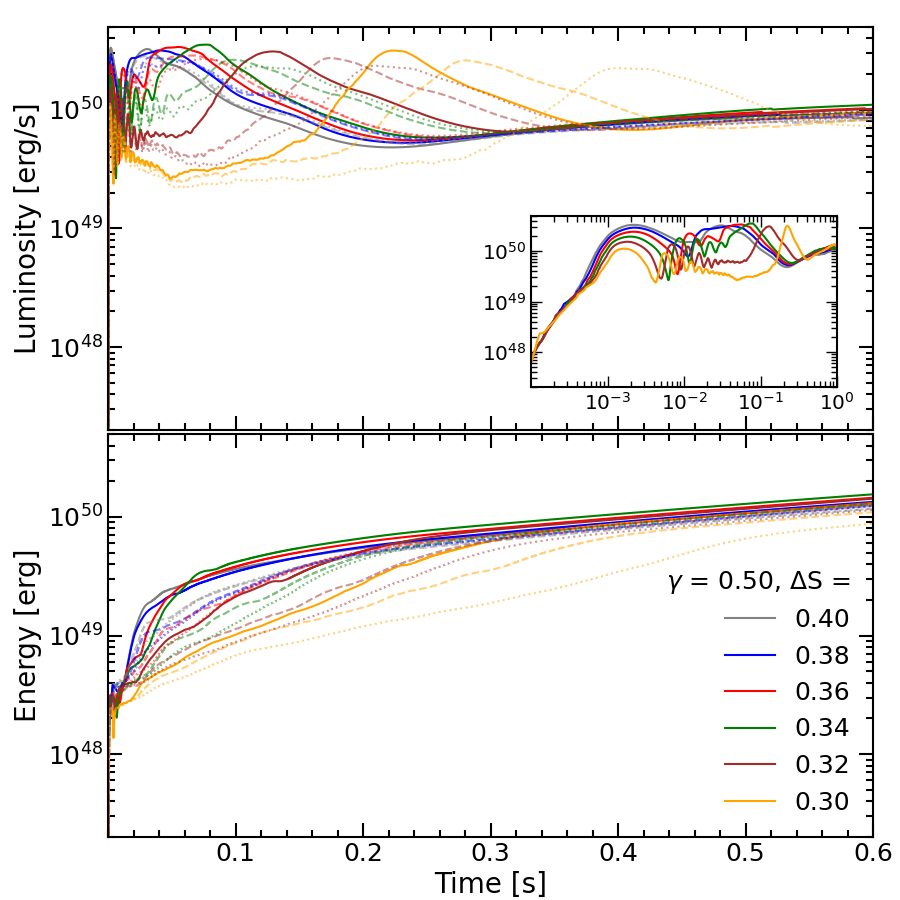}
    \caption{
    Same as Figure~\ref{fig:network energy among gamma}, but for the models L-40-50, L-38-50, L-36-50, L-34-50, L-32-50, and L-30-50. 
    The nuclear reaction network is updated for every $\delta t_{\text{net}} =$ $1\times10^{-4}$\,s (solid lines), $2\times10^{-4}$\,s (dashed lines), and $5\times10^{-4}$\,s (dotted lines) before the explosions.
    }
    \label{fig:network energy among delta S}
\end{figure}

The evolution of the central density, electron antineutrino luminosity, and net cumulative thermal energy deposited on the neutron star by the network in the models L-40-50, L-38-50, L-36-50, L-34-50, L-32-50, and L-30-50 are shown in Figures~\ref{fig:central density among delta S} and \ref{fig:network energy among delta S}. 
As a consequence of the difference in explosion timescale discussed in Section~\ref{sec:Explosion of subminimal neutron star}, the neutrino luminosity of models with lower $\Delta S$ fluctuates periodically following the radial oscillations of the star prior to the explosion. 
A peak luminosity of $\sim3\times10^{50}$\ergs\, is always found in these models when the explosion occurs. 
The net cumulative thermal energy deposited on the star by the network is similar, starting from $\sim0.3$\,s after the occurrence of explosion in all these models.

%----------------------------------------------------------------------
\section{Numerical test}
\label{app: numerical test}

Owing to limited computational resources, we do not solve the nuclear reaction network for every time step determined by the CFL condition during the hydrodynamic simulations. 
The extra energy source terms in Equation~(\ref{eq:energy equation}) are updated by solving the network with time steps $\delta t_{\text{net}}$ shorter than the $\beta$-decay half-lives $\sim10^{-3}$\,s of the neutron-rich nuclei presented. 
In Figures~\ref{fig:central density among delta S} and \ref{fig:network energy among delta S}, we update the network for every $\delta t_{\text{net}} =$ $1\times10^{-4}$\,s (solid lines), $2\times10^{-4}$\,s (dashed lines), and $5\times10^{-4}$\,s (dotted lines) before the explosions. 
We notice that the explosions are advanced in all models whenever a smaller $\delta t_{\text{net}}$ is picked as we approach the physical limit $\delta t_{\text{net}} = 0$. 
In particular, the explosion timescale as well as the peak in neutrino emission have not fully converged even for the finest $\delta t_{\text{net}}$ adopted.

Despite the difference in the hydrodynamics, the overall and local net energy generated by the network are insensitive to the choice of $\delta t_{\text{net}}$. 
Moreover, the features of chemical element productions discussed in Section~\ref{sec:Nucleosynthesis} are robust against the value of $\delta t_{\text{net}}$ chosen. 
The analysis performed in this paper is based on the simulation results using $\delta t_{\text{net}} =$ $1\times10^{-4}$\,s. 
Further investigation is required to study the exact explosion time of subminimal mass neutron star explosions.

\bibliography{sample631}{}
\bibliographystyle{aasjournal}

\end{document}